\documentclass[aps,10pt,twocolumn,prd,floats,floatfix]{revtex4-1}
\usepackage{verbatim}
\usepackage{listings}
\usepackage{graphicx}
\usepackage{color}
\usepackage{amsmath}
\usepackage{amssymb}
\usepackage[dvips]{epsfig}
\usepackage[T1]{fontenc}
\usepackage{shadow}
\usepackage{varioref}
\usepackage{epsfig}
\usepackage{array}
\usepackage{rotating}

\usepackage{myaasmacros}

\newcommand{\Sgra}{Sgr~A$^*$}
\newcommand{\pos}{$\beta^+$}
\newcommand{\mr}{\mathrm}

\newcommand{\be}{\begin{equation}}
\newcommand{\ee}{\end{equation}}
\newcommand{\bea}{\begin{eqnarray}}
\newcommand{\eea}{\end{eqnarray}}

\begin{document}
%\title{A new Lyman-$\alpha$ limit on extragalactic magnetic fields}
\title{Positron excess in the center of the Milky Way from short-lived $\beta^+$ emitting isotopes }

\author{M. S. Pshirkov$^{1,2,3}$\\
$^{1}$ {\footnotesize{\it Sternberg Astronomical Institute, Lomonosov Moscow State University, Universitetsky prospekt 13, 119992, Moscow, Russia}}\\
$^{2}$ {\footnotesize{\it Institute for Nuclear Research of the Russian Academy of Sciences, 117312, Moscow, Russia}}\\
$^{3}$ {\footnotesize{\it Pushchino Radio Astronomy Observatory, 142290 Pushchino, Russia}}\\
}
\noaffiliation

\begin{abstract} 
Observations of the \textit{INTEGRAL} satellite revealed the presence of yet unexplained excess in the central region of the Galaxy at  the energies around 511 keV. These gamma-rays are produced in the process of positron annihilation,  the needed  rate is around  $10^{42}~\mathrm{s^{-1}}$. In this short paper it is shown that  \pos -emitting isotopes that are formed in interactions of subrelativistic cosmic rays (CRs) with light nuclei (CNONe) can account for a considerable fraction -- up to several  tens of percent -- of $e^{+}$ production rate in the central region. 
\end{abstract}

\maketitle

%%%%%%%%%%%%%%%%%%%%%%%%%%%%%%%%%%%%%%%%%%%%%%%%%%%%%%%
%\section{Introduction}\label{sec:mot}
%%%%%%%%%%%%%%%%%%%%%%%%%%%%%%%%%%%%%%%%%%%%%%%%%%%%%%%
\textbf{Introduction.} The central region  is the most mysterious part of our Galaxy and  hosts a lot of different astrophysical phenomena. One of them  
is an  excess in the 511~keV $\gamma$-line that  was observed by the SPI/\textit{INTEGRAL} instrument \cite{INTEGRAL1,INTEGRAL2, INTEGRAL3}. This excess can be naturally attributed to the positron annihilation in this region. Recent analysis   confirmed existence of the central source at the statistical significance $\sim5\sigma$ and its corresponding steady-state production rate $\dot{N}_{e^+}=(0.3-1.2)\times 10^{42}$~s$^{-1}$ \cite{Siegert2016}. Because of limited angular resolution of the instrument there is only an upper limit on the size of this source: $r<2.7^{\circ}$, so it could be either point-like, possibly connected with a super-massive black hole Sgr A$^*$ in the very center of our Galaxy or it could span larger region up to several hundred pc size.    
A large number of candidates  were proposed in order to explain this excess: $e^+$ could be produced in nucleosynthesis processes related to supernovae explosions --  decays of  radioactive nuclei $^{26}$Al, $^{44}$Ti, $^{56}$Ni \cite{Prantzos2006,Alexis2014}, they could be alternatively produced in the immediate vicinity of the Sgr A$^*$ \cite{Totani2006} or in numerous microquasars \cite{Siegert2016b}. Positron production could be non-stationary, taking place during high-states of the \Sgra activity or star-bursts \cite{Parizot2005,Totani2006,Cheng2006}. Finally, there are also  large number of dark matter  models, which could explain the excess, see e.g. \cite{Dubovsky2004,Ascasibar2006,Finkbeiner2007, Pospelov2007}. A broad spectrum of possible production mechanisms is reviewed in \cite{Prantzos2011}.

In this short paper it is shown that \pos -emitting isotopes that are formed in interactions of subrelativistic cosmic rays (CRs) with CNONe atoms can account for a considerable fraction of $e^{+}$ production rate in the central region. The idea is simple: the region demonstrates a high uniform degree of ionization that can be caused by CRs permeating it \cite{Dogiel2015,Nobukawa2015}. Alternative viable explanation is heating by turbulent motions \cite{Ginsburg2016}, or, naturally, both processes could contribute to ionization simultaneously.

High-energy observations at energies larger than 100~MeV by the Fermi LAT instrument constrain these CRs to be mostly sub-relativistic with energy density in the central hundred pc around $50-80~\mathrm{eV~cm^{-3}}$ \cite{Dogiel2015}. If their spectrum is hard, then flux at $E=200$~MeV could be as high as $\phi_{CR}=5\times10^2~\mathrm{cm^{-2}~s^{-1}~sr^{-1}}$, and this value will be used as a benchmark.
The central region contains a considerable fraction of the total mass of the galactic gas in the so-called Central Molecular Zone (CMZ) --  with a radius 200-300 pc and height $\sim$100 pc, its mass can reach $7\times 10^{7}~M_{\odot}$ \cite{Morris1996,Launhardt2002}. There are two main components  in the CMZ:  dense clumps with densities $\sim10^{4}~\mathrm{cm^{-3}}$ and surrounding more tenuous medium  ($\sim10^{2}~\mathrm{cm^{-3}}$).

\textbf{Method.} Radioactive \pos isotopes could be produced in abundance when CRs are interacting with atoms of light elements, such as C, N, O, Ne. These proton-induced reactions have low thresholds around 10-20~MeV and  produce positrons with energies 1-2~MeV. Their cross-section values were taken from the JENDL library  \cite{JANIS,JENDL} with the only exception of the $^{21}Ne+p \rightarrow ^{18}F +\alpha$ process, which cross-section was estimated using TENDL library \cite{TENDL}:
\begin{align}
 ^{12}C+p \rightarrow ^{11}C +d,~ ^{11}C \rightarrow ^{11}B + e^{+}, \\ \nonumber T_{1/2}\sim 1200 ~\mathrm{s}, E=0.97~\mathrm{MeV},~ \sigma_{200~{\mathrm{MeV},~}}=40~\mathrm{mb}\\  
   ^{14}N+p \rightarrow ^{14}O +n,~ ^{14}O \rightarrow^{14}N + e^{+}, \\ \nonumber T_{1/2}\sim 70 ~\mathrm{s}, E=0.78~\mathrm{MeV},~ \sigma_{200~{\mathrm{MeV},~}}=7.5~\mathrm{mb}\\ 
   ^{14}N+p \rightarrow ^{13}N +d,~ ^{13}N \rightarrow^{13}C + e^{+},\\ \nonumber T_{1/2}\sim 600 ~\mathrm{s}, E=1.19~\mathrm{MeV},~ \sigma_{200~{\mathrm{MeV},~}}=9~\mathrm{mb}\\  
 ^{14}N+p \rightarrow ^{11}C +\alpha, \\ \nonumber T_{1/2}\sim 1200 ~\mathrm{s}, E=0.97~\mathrm{MeV},~
   \sigma_{200~{\mathrm{MeV},~}}=24~\mathrm{mb}\\  
  ^{16}O+p \rightarrow ^{15}O +d,  ^{15}O \rightarrow ^{15}N + e^{+} \\ \nonumber T_{1/2}\sim 122 ~\mathrm{s}, E=1.7~\mathrm{MeV},~ \sigma_{200~{\mathrm{MeV},~}}=40~\mathrm{mb}\\   
  ^{16}O+p \rightarrow ^{11}C +n+p+\alpha,   \\ \nonumber T_{1/2}\sim 1200 ~\mathrm{s}, E=0.97~\mathrm{MeV},~ \sigma_{200~{\mathrm{MeV},~}}=10~\mathrm{mb} \\
  ^{16}O+p \rightarrow ^{13}N +\alpha,   \\ \nonumber T_{1/2}\sim 600 ~\mathrm{s}, E=1.19~\mathrm{MeV},~ \sigma_{200~{\mathrm{MeV},~}}=7.5~\mathrm{mb}\\  
  ^{21}Ne+p \rightarrow ^{18}F +\alpha, ^{18}F \rightarrow^{18}O+ e^{+} \\ \nonumber T_{1/2}\sim 6600 ~\mathrm{s}, E=0.64~\mathrm{MeV},~ \sigma_{200~{\mathrm{MeV},~}}=6~\mathrm{mb}
  \end{align}

  Relevant cross-sections can be written out for each nuclei species :
  \begin{eqnarray}
\sigma_C(\mathrm{200~MeV})=40~\mr{mb} \\ 
\nonumber \sigma_N(\mathrm{200~MeV})=40~\mr{mb} \\
\nonumber \sigma_O(\mathrm{200~MeV})=60~\mr{mb} \\
\nonumber \sigma_{Ne}(\mathrm{200~MeV})=6~\mr{mb}
\label{eq:cross-sections}
  \end{eqnarray}

  Production rate  can be estimated as follows:
  
  \begin{equation}
   \label{eq:prod-rate}
   \mathcal{R}=4\pi \int \phi_{CR} \sum\limits_{i} n_i\sigma_i dV = 4\pi \phi_{CR} \sum\limits_{i} N_i\sigma_i,
  \end{equation}
  %It is assume that the cosmic ray flux $\phi_{CR}$ level is constant across the entire CMZ 
where $\phi_{CR}$ is a CR flux, which level is adopted to be constant across the central region \cite{Dogiel2009}, $n_i, \sigma_i, N_i$ are number density, cross-section (see  Eq. (\ref{eq:cross-sections})), and total number of $i$-species atoms in the region, correspondingly.
Total number of atoms $N_i$ can be linked with number of H atoms in the CMZ and, eventually, with its mass:
  \begin{equation}\label{eq:atom_number}
   N_i=K \eta 10^{X_{i\odot}-12.0}M_H/m_p,
  \end{equation}
  where a coefficient $K=2-5$ describes  enhancement of metallicity in the center \cite{Daflon2004,Rudolph2006}, $X_{i\odot}$ is the solar system abundance of $i$-th species \cite{Lodders2010}, $M_H$ is the total mass of the hydrogen in the CMZ, $M_H=0.75M_{CMZ}$, $m_p$ is the proton mass, and $\eta \leq 1$ describes the suppression effect that arises due to an inability of low-energy CRs to penetrate inside dense clouds \cite{Dogiel2015}. On the other hand, this impediment could lead to enhanced rate of interactions in the boundary/envelope regions of dense clouds. The morphological  properties of these clouds are highly uncertain and all the complexity of effects involved is encoded in a single coefficient $\eta$. In order to get an upper limit on possible positron production, values $K=3,~\eta=1, M_H=5\times10^{7}~M_{\odot}$ were adopted in the subsequent calculations.   
  
  Number of nuclei can be readily estimated using solar abundances $X_{C\odot}=8.39$, $X_{N\odot}=7.86$, $X_{O\odot}=8.73$, $X_{Ne\odot}=8.05$ \cite{Lodders2010}:
      \begin{eqnarray}\label{eq:number_atoms_CMZ}
N_C=1.5\times10^{61}K \\ 
\nonumber N_N=4.5\times10^{60}K \\
\nonumber N_O=3.4\times10^{61}K \\
\nonumber N_{Ne}=7.0\times10^{60}K
\end{eqnarray}
  
  Combining (\ref{eq:cross-sections}) with (\ref{eq:number_atoms_CMZ}) we obtain  the final expression:
  
    \begin{align}\label{eq:prod-rate2}
   \mathcal{R}=5.4\times 10^{40} \eta \left(\frac{K}{3}\right)\left(\frac{\phi_{CR}}{5\times10^2~\mathrm{cm^{-2}~s^{-1}~sr^{-1}}}\right)\times \\ \nonumber\times\left(\frac{M_{200~\mr{pc}}}{7\times 10^{7}~M_{\odot}}\right)~\mr{s^{-1}},
  \end{align}
  where $M_{200~\mr{pc}}$ is the total gas mass in $r=200$~pc radius.
 There also could be some considerable contribution from reverse spallation process, i.e. when CNONe CRs hit $p-$ or He- targets.  The rate calculated above can significantly contribute to the total production rate $(3 - 12)\times10^{41}~\mr{s^{-1}}$ that was estimated from the observations \cite{Siegert2016}.
 
 The shape of the CR spectrum is very important -- all previous considerations used a very simplistic approach of a mono-energetic spectrum, which in fact is a good approximation for a very hard spectrum --  $dN/dE\propto E^{-0.5}$   \cite{Dogiel2015}. If instead much softer spectrum, like $dN/dE\propto E^{-2.5}$ \cite{Nobukawa2015} was used, then things get  more complicated: it is  necessary to convolve the spectrum with individual energy-dependent cross-sections \cite{JANIS}  down to the thresholds of corresponding reactions  of Eq. (\ref{eq:cross-sections}). 
 We use the spectrum from the \cite{Nobukawa2015} paper: $dN/dE=A_0 (E/E_0)^{-2.5}$, $A_0=1.4\times10^{-11}~\mr{MeV^{-1}~cm^{-3}}$ is the normalization coefficient at the energy scale  $E_0=1$~MeV. 'Effective' cross-sections  can  substitute for cross-sections given in Eq. (\ref{eq:cross-sections}):
 
     \begin{equation}\label{eq:cross-sections_eff}
\sigma_{eff}=\frac{\int_{E_{\mr{min}}}^{E_{\mr{max}}}\sigma(E)dN/dE(E)v(E)dE}{4\pi\phi_{CR}}
  \end{equation}.

    \begin{eqnarray}
\sigma_{effC}=10~\mr{mb},\\ 
\nonumber \sigma_{effN}=90~\mr{mb},\\
\nonumber \sigma_{effO}=18~\mr{mb},\\
\nonumber \sigma_{effNe}=36~\mr{mb},
\label{eq:cross-sections_eff2}
  \end{eqnarray}
  and the corresponding luminosity:
    \begin{align}
    \label{eq:prod-rate3}
   \mathcal{R}_{\mr{soft}}=2.7\times 10^{40} \eta \left(\frac{K}{3}\right)\left(\frac{\phi_{CR}}{5\times10^2~\mathrm{cm^{-2}~s^{-1}~sr^{-1}}}\right)\times\\ \nonumber\times\left(\frac{M_{200~\mr{pc}}}{7\times 10^{7}~M_{\odot}}\right)~\mr{s^{-1}},
  \end{align}
  which is only two times smaller than value from Eq. (\ref{eq:prod-rate2}) despite large difference between the respective spectra.  Obviously, if the spectrum demonstrated a break around 10 MeV, the luminosity can be even higher than estimated in Eq. (\ref{eq:prod-rate2}).

  \textbf{Related phenomena.} 
  Collisions of subrelativistic cosmic rays with CNO  can also result in an emergence of nuclear gamma-ray lines: excited nuclei during transition   to their lower energy levels radiate photons  with characteristic energies around several MeV  \cite{Dogiel2009b}. E.g., one of the most promising target is the 4.44 MeV line from the deexcitation of $^{12}C^*$  nuclei, which in turn can be produced in inelastic collisions  with protons and also in processes of proton-induced spallation of N and O nuclei. The resulting flux at the Earth can be  calculated using cross-sections from \cite{Kozlovsky2002}. In case of very hard spectrum ($dN/dE\propto E^{-0.5}, E_{\mr{max}}=200$~MeV) the  flux is equal to:
  \begin{equation}
   \label{gamma_flux_1}
   F_{4.44~MeV}\sim10^{-6}(K/3)~\mr{ph~cm^{-2}~s^{-1}},
  \end{equation}
and it increases sixfold up to $6\times10^{-6}(K/3)~\mr{ph~cm^{-2}~s^{-1}}$ in case of soft spectrum ($dN/dE\propto E^{-2.5}$). Unfortunately, in both cases it is below the sensitivity threshold of the SPI/\textit{INTEGRAL} instrument. However, future missions such as proposed \textit{GRIPS} \cite{GRIPS} can be sensitive enough to detect these elusive lines that come from p-CNO interactions.
  
%   Another phenomenon that can be related  with the proposed mechanism is the de-excitation gamma-ray lines that are also  produced in the very same $p$-C, $p$-N, and $p$-O collisions and have energies around $\sim5$~MeV \cite{Dogiel2009b}. It is easy to link these processes  in the case of quasi mono-energetic spectrum peaking at $E>100$~MeV \cite{Murphy2009}: $R_{\gamma}/R_{e^+}=\sigma_{\gamma}/\sigma_{e^+}=0.2$. 
%  Relation will be  less clear-cut in the  case of softer spectrum.

  The  $^{11}C$  isotope forms in several reactions and afterwards quickly  decays into stable  $^{11}B$ with the  production rate $\mathcal{R}_B$ that is equal to almost  $40\%$ of total rate $\mathcal{R}$. In 10 Gya  it would lead to local enrichment which could be roughly estimated neglecting advection and changes in CMZ composition and CR flux levels throughout galactic history: $X_B=5.0$. This value  is much higher than $X_{B\odot}=2.8$.  Unfortunately, boron searches require UV-observations that can hardly be performed for  the very central region of the Galaxy. This boron overabundance can also manifest itself in alteration of the observed local B/C ratio if boron nuclei, especially produced in the spallation scenario,  were accelerated and eventually propagated into the Outer Galaxy. Also spectral studies of hyper-velocity stars which could originate from the very vicinity of the GC \cite{Hills1988} can be used for testing the model -- they can demonstrate unusually high levels of $^{11}B$, though this characteristic can also be erased during star's evolution \cite{Przybilla2008}.

  \textbf{Conclusions.} Short-lived \pos -emitting isotopes can be produced in interactions of subrelativistic cosmic rays (CRs) with light nuclei in the CMZ. In case of very hard spectrum of the CRs, $dN/dE\propto E^{-0.5}$, the positron production rate can be as high as $5.4\times 10^{40}~\mr{s^{-1}}$, while the production is somewhat suppressed for softer spectra, $\mathcal{R}=2.7\times 10^{40}~\mr{s^{-1}}$ for  $dN/dE\propto E^{-2.5}$. It can account for up to 20\% of total positron production rate in the central region. The corresponding gamma-ray emission from nuclear de-excitation lines is too weak to be observed with current instruments but can be detected with future MeV-detectors.

%%%%%%%%%%%%%%%%%%%%%%%%%%%%%%%%%%%%%%%%%%%%%%%%%%%%%%%
% \section{Method}\label{sec:met}
%%
%%%%%%%%%%%%%%%%%%%%%%%%%%%%%%%%%%%%%%%%%%%%%%%%%%%%%%%
% \section{Discussion}\label{sec:dis}
%%%%%%%%%%%%%%%%%%%%%%%%%%%%%%%%%%%%%%%%%%%%%%%%%%%%%%%

%%%%%%%%%%%%%%%%%%%%%%%%%%%%%%%%%%%%%%%%%%%%%%%%%%%%%%%
\paragraph*{Acknowledgements.}  The author would like  to thank K. Postnov and D. Chernyshov for insightful discussions and G.Rubtsov and S.Troitsky for valuable comments on the manuscript. The work of the author was supported by  the Russian Science Foundation grant 14-12-01340. This research has made use of NASA's Astrophysics Data System.

\bibliography{positrons}

\end{document}